\documentclass[reprint,amsmath,amssymb,aps,prl,groupedaddress,nofootinbib,twocolumn,superscriptaddress]{revtex4-1}

\usepackage{graphicx}
\usepackage{amsthm,amssymb,amsmath,braket,mathdots}
\usepackage{bm}
\usepackage[pagebackref=false,pdfnewwindow=true]{hyperref} %\hypersetup{draft}
\usepackage{epstopdf,psfrag}
\usepackage{relsize,amsbsy}
\usepackage[export]{adjustbox}
\usepackage{makecell}
\usepackage{graphicx,xcolor,tikz}
\usepackage{float}
\usepackage[centerlast]{caption}
\usepackage{hyperref}

\newcommand{\be}{\begin{equation}}
\newcommand{\ee}{\end{equation}}

\newcommand{\bit}{\begin{enumerate}}
	\newcommand{\eit}{\end{enumerate}}

\definecolor{bananayellow}{rgb}{1.0, 0.88, 0.21}
\definecolor{straw}{rgb}{0.32, 0.28, 0.1}

\begin{document}
	\title{Orthogonal Quantum Many-body Scars}
		\author{Hongzheng Zhao} 
		\affiliation{\small Blackett Laboratory, Imperial College London, London SW7 2AZ, United Kingdom}
		\author{Adam Smith}
		\affiliation{School of Physics and Astronomy, University of Nottingham,
University Park, Nottingham NG7 2RD, United Kingdom}
		\author{Florian Mintert}
	\affiliation{\small Blackett Laboratory, Imperial College London, London SW7 2AZ, United Kingdom}
	
		\author{Johannes Knolle  }
	\affiliation{Department of Physics TQM, Technische Universit{\"a}t M{\"u}nchen, James-Franck-Stra{\ss}e 1, D-85748 Garching, Germany}
	\affiliation{Munich Center for Quantum Science and Technology (MCQST), 80799 Munich, Germany}
	\affiliation{\small Blackett Laboratory, Imperial College London, London SW7 2AZ, United Kingdom}
	
	\begin{abstract}
	Quantum many-body scars have been put forward as counterexamples to the Eigenstate Thermalization Hypothesis. These atypical states are observed in a range of correlated models as long-lived oscillations of local observables in quench experiments starting from selected initial states. The long-time memory is a manifestation of quantum non-ergodicity generally linked to a sub-extensive generation of entanglement entropy, the latter of which is widely used as a diagnostic for identifying quantum many-body scars numerically as low entanglement outliers. Here we show that, by adding kinetic constraints to a fractionalized orthogonal metal, we can construct a minimal model with {\it orthogonal} quantum many-body scars leading to persistent oscillations with infinite lifetime coexisting with rapid volume-law entanglement generation. Our example
    provides new insights into the link between quantum ergodicity and many-body entanglement while opening new avenues for exotic non-equilibrium dynamics in strongly correlated multi-component quantum systems.

	\end{abstract}

	\maketitle

\textit{Introduction.--}
Rapid experimental progress in state preparation and controlability, e.g., in cold-atoms ~\cite{yang2020observation,kaufman2016quantum,scherg2020observing,choi2016exploring,zhang2017observation,bernien2017probing,lukin2019probing}, ion-traps~\cite{martinez2016real,brydges2019probing} or superconducting circuits~\cite{neill2016ergodic,xu2018emulating}, have  enabled the exploration of non-equilibrium dynamics of quantum many-body systems in well-isolated settings. The Eigenstate Thermalization Hypothesis (ETH) provides a basic framework for understanding how these systems approach thermal equilibrium under unitary time-evolution~\cite{deutsch2018eigenstate,srednicki1994chaos,rigol2008thermalization,cassidy2011generalized}. It states that local observables calculated in  generic eigenstates look thermal, i.e. only depend on the eigenenergy. Consequently, in a quantum quench protocol the memory of  initial states is quickly lost. This ergodicity is also manifest in the extensive (volume-law) scaling of the entanglement entropy of eigenstates as a function of sub-system size as confirmed by numerous numerical studies~\cite{d2016quantum}. Recently, the research focus shifted to systems which deviate from the ETH paradigm~\cite{shiraishi2017systematic}, for instance integrable systems~\cite{rigol2009breakdown,rigol2007relaxation} or disordered many-body localized phases (MBL)~\cite{abanin2017recent,pal2010many, abanin2019colloquium}. These systems are capable of retaining certain memory of initial states as observed in quench experiments~\cite{choi2016exploring,kohlert2019observation,schreiber2015observation,smith2016many,lukin2019probing,xu2018emulating}. This non-ergodicity is also manifest in the non-extensive (area-law) entanglement scaling of MBL eigenstates~\cite{abanin2019colloquium}.

It came as a surprise when a Rydberg atom experiment reported the breaking of ergodicity in a clean and non-integrable system as observed  via long-lived oscillations after quenching from specific initial states~\cite{bernien2017probing}. Subsequently, it was proposed that it is the kinetic constraints which lead to special eigenstates dubbed {\it quantum many-body scars}~\cite{moudgalya2018exact,moudgalya2018entanglement,turner2018weak,choi2019emergent,bull2019systematic,ho2019periodic,iadecola2019quantum,pai2019dynamical,schecter2019weak,moudgalya2020eta,iadecola2020quantum,bluvstein2020controlling}. These ETH-violating states may appear throughout an otherwise ETH-obeying spectrum and perfect scarred states populate only a small isolated sector of the entire many-body Hilbert space~\cite{sala2020ergodicity}. 

In general, quantum many-body scars are expected to show sub-extensive entanglement scaling, which indeed is routinely used to detected them numerically as entropy-outliers~\cite{serbyn2020quantum}. In the Rydberg experiment, simple product states were chosen as initial states which have substantial overlap with the scar states, leading to persistent oscillations and low entanglement generation~\cite{turner2018weak}. 

Here, we address the intriguing question of whether long-lived coherent oscillations can coexist with rapid volume-law entanglement generation in a standard quench-setup of a clean non-integrable quantum many-body system? 
The intuitive answer should be {\it No} because within the ETH (or generalized ETH for integrable systems \cite{cassidy2011generalized,rigol2007relaxation}) paradigm volume-law entanglement normally goes hand-in-hand with thermal relaxation of local observables in static quantum many body systems. Similarly, the fact that proposals to date show the coexistence of persistent oscillations with area-law entanglement entropy~\cite{serbyn2020quantum} has been thought to be dictated by a much more general underlying structure of quantum thermalization.
However, in strongly correlated quantum many-body systems, ETH can be violated in unexpected ways and we answer the above question affirmatively by constructing a concrete counter-example.

Multi-component systems have been suggested to evade thermalization via inter-species interactions, for example in heavy-light particle mixtures~\cite{kagan1984localization,papic2015many,yao2016quasi}. The original idea is that even in the absence of quenched disorder, the light particles localize on the disordered background of slow heavy particles which themselves can be in thermal equilibrium. Subsequently, a general class of non-ergodic phases known as quantum disentangled liquids (QDLs) were proposed~\cite{grover2014quantum}, in which some degrees of freedom (d.o.f.) exhibit area-law entanglement, while others are volume-law entangled. QDL-like behavior has by now been identified in a range of correlated systems, e.g., in lattice gauge models~\cite{smith2017disorder,smith2018dynamical,brenes2018many,karpov2020disorder,paulson2020towards,papaefstathiou2020disorder}, the half-filled Hubbard model~\cite{veness2017quantum} and frustrated quantum magnets~\cite{mcclarty2020disorder,zhu2020subdiffusive}, but  none provide an answer to our main question. 

We will provide a basic construction of a QDL showing infinitely long-lived persistent oscillations and rapid volume-law entanglement generation. Our starting point is the Orthogonal Metal (OM) model of Ref.~\cite{nandkishore2012orthogonal}. OMs were introduced as a particularly simple example of a non-Fermi liquid in which the original physical d.o.f. split into separate components, each carrying fractions of the original quantum numbers.
These components have their own dynamics in sharp contrast to the emergent static gauge d.o.f. in, {\it e.g.} Kiteav model \cite{kitaev2003fault} or lattice gauge models exhibiting disorder-free localization \cite{smith2018dynamical}. OMs were named {\it orthogonal} because basic spectral functions of the physical d.o.f. can be zero despite the presence of a well defined Fermi sea and non-zero conductivity originating from one of the components. We will show how to combine the fractionalization mechanism of the OM with kinetically constrained tunneling~\cite{zhao2020quantum,hudomal2020quantum,de2019dynamics,brighi2020stability}, which results in Hilbert space fragmentation \cite{sala2020ergodicity,khemani2020localization} in one component, for realizing {\it orthogonal} quantum many-body scars.

\textit{The Model.--}
Our starting point is the following one-dimensional Hamiltonian with two species of particles
\begin{eqnarray}
\begin{aligned}
\label{eq.original}
H =-\sum_{i}\Big(h \sigma_{i-1, i}^{x} \sigma_{i, i+1}^{x}(-1)^{c_{i}^{\dagger} c_{i}}+g_z\sigma_{i, i+1}^z\sigma_{i+1, i+2}^z \\+J \sigma_{i, i+1}^{z}\Big)-t \sum_{i}\left(n_{i-1}c_{i}^{\dagger} \sigma_{i,i+1}^z c_{i+1}n_{i+2}+ \text{H.c.}\right).
\end{aligned}
\end{eqnarray}
The $c_i^{\dagger}$ operators create physical spinless fermion d.o.f., at site $i$, and the $\sigma_{i,i+1}$ operators represent a spins-$\frac{1}{2}$ background field positioned at the links between site $i$ and $i+1$ of a one dimensional lattice. A similar model with a standard unconstrained fermionic hopping and vanishing interaction constant $g_z$ had been  introduced as a basic solvable example of the OM  \cite{nandkishore2012orthogonal}. We will show that the addition of  the density dependence of hopping processes introduces kinetic constraints~\cite{zhao2020quantum,hudomal2020quantum} which is key for inducing non-ergodicity. 
 
While the original d.o.f. are strongly coupled, we can show explicitly how the different sectors emerge.  
First, a duality transformation~\cite{kramers1941statistics,wegner1971duality} maps the link-spins $\sigma$ to a site-spin $\tau$ as $\tau_i^z=\sigma_{i-1, i}^{x} \sigma_{i, i+1}^{x},\  \tau_i^x\tau_{i+1}^x=\sigma_{i,i+1}^z,$ such that the Hamiltonian reduces to
\begin{eqnarray}
\begin{aligned}
H =&-\sum_{i}\left(J \tau_{i}^{x} \tau_{i+1}^{x}+h \tau_{i}^{z}(-1)^{c_{i}^{\dagger} c_{i}}+g_z{\tau}^x_i{\tau}^x_{i+2}\right) 
\\&-t \sum_{i}\left(n_{i-1}c_{i}^{\dagger} \tau_{i}^{x} \tau_{i+1}^{x} c_{i+1}n_{n+2}+ \text{H.c.}\right).
\end{aligned}
\end{eqnarray}
Next, we can define new composite d.o.f. for fermions  $f_{r}=\tau_{r}^{x} c_{r}$ and dual spins $\tilde{\tau}_{i}^{z}=\tau_{i}^{z}(-1)^{f_{i}^{\dagger} f_{i}}, \quad \tilde{\tau}_{i}^{x}=\tau_{i}^{x}$.
Finally, in the new variables the Hamiltonian separates into two components $H=H_{\tilde{\tau}}+H_f$
\begin{eqnarray}
\begin{aligned}
H_{\tilde{\tau}}&=-\sum_{ i} \left( J\tilde{\tau}_{i}^{x} \tilde{\tau}_{i+1}^{x}+h  \tilde{\tau}_{i}^{z}+g_z\tilde{\tau}_{i}^{x}\tilde{\tau}_{i+2}^{x}\right),\\
H_{f} &=-t \sum_{i}\left(n_{i-1}f_{i}^{\dagger} f_{i+1}n_{i+2}+ \text{H.c.}\right),
\end{aligned}
\end{eqnarray} 
a constrained $f-$fermion system and a non-integrable $\tilde{\tau}-$spin chain. A crucial point is that the dynamics of the physical variables is a combination of these separate dynamical sectors. We note, that one can introduce different types of interactions while retaining the separability. For instance, since the fermionic particle number $n_i$ are the same in both $c-$ and $f-$representations, the density-density interaction $\sum_{i,j}n_in_j$ can be introduced meanwhile $\tilde{\tau}$ remains untouched.

\begin{figure}
	\centering
	\includegraphics[width=0.99\linewidth]{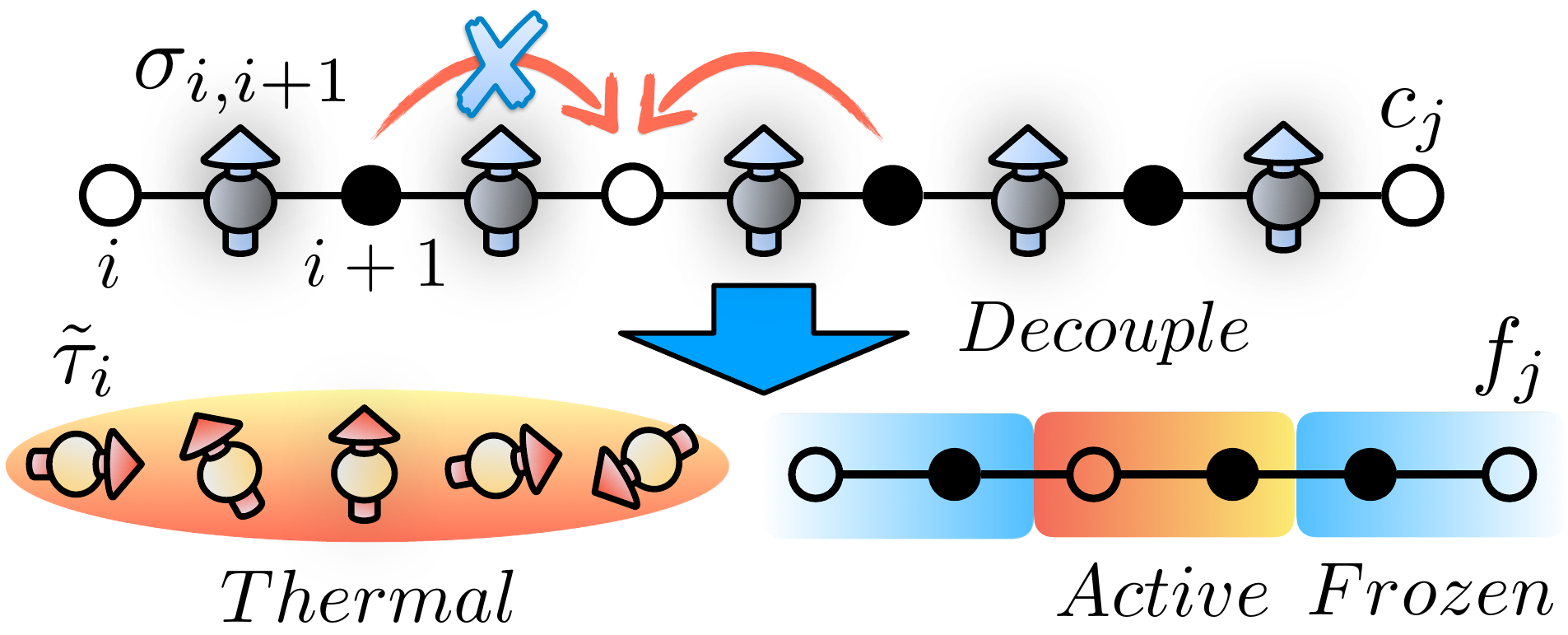}
	\caption{Schematic picture of our multi-component model, with $\sigma-$spins living on the links between $c-$fermions. Filled black circles denote occupied fermionic sites. The dynamics for this particular initial state is kinetically constrained. The model can be decoupled into dual $\tilde{\tau}-$spins and $f-$fermion. The dynamics for the latter is only allowed on active sites hence hosting persistent coherent dynamics, while the $\tilde{\tau}-$spins thermalize. }
	\label{fig:model}
\end{figure}

The density-dependent hopping processes in $H_f$ significantly restrict the dynamics of $f$-fermions, i.e. hopping is only permitted when two sites are both surrounded by occupied ones. Such kinetic constraints result in non-thermal isolated fragments in the Hilbert space ~\cite{sala2020ergodicity,khemani2020localization} thus fermionic dynamics can now be non-ergodic for properly chosen initial states as elaborated in the following example.
Consider the Fock state of six fermionic sites $\ket{010110}_{{f}}$, hopping is only allowed in the middle two sites as
\begin{eqnarray}
\label{eq.twodimension}
H_f^2\ket{010110}_f = -H_ft\ket{011010}_f=t^2\ket{010110}_f\ .
\end{eqnarray}
All other sites are frozen, thus reducing the effective Hilbert space dimension to 2. Crucially, this state can be further treated as a building block to form states of larger sizes at a finite fermion density, for example the product of the same state $\dots\ket{010110}_f\otimes\ket{010110}_f\dots$. As there is no mixing allowed between different blocks, the tensor product is retained during time evolution, and the dimension of the reduced space, $2^{L/6}$ for a chain of length $L$, is substantially smaller than the total Hilbert space \cite{zhao2020quantum,hudomal2020quantum}. Therefore, starting with this initial state, the dynamics of a quench governed by the constrained model $H_f$ will only appear in small separated subspaces which results in persistent coherent oscillations.

On the contrary, the spin Hamiltonian $H_{\tilde{\tau}}$ reduces to the transverse field Ising model~\cite{pfeuty1970one} for vinishing $g_z$, but we chose a finite value to break its integrability such that local observables rapidly approach a stationary state of thermal equilibrium together with volume-law entanglement~\cite{calabrese2005evolution}.

The dynamics of the physical d.o.f. Eq.~\eqref{eq.original} can now be understood from the decoupling scheme. For instance, the physical correlation function and magnetization for the background $\sigma-$spin are given by
\begin{eqnarray}
\begin{aligned}
\label{eq.spin}
\langle \sigma_{i-1, i}^x \sigma_{i, i+1}^x\rangle = \langle \tilde{\tau}_i^z\rangle (1-2 \langle n_i \rangle),\hspace{1pc} \langle \sigma_{i,i+1}^z\rangle=\langle \tilde{\tau}_i^x\tilde{\tau}_{i+1}^x\rangle. \ \ \ 
\end{aligned}
\end{eqnarray}
Crucially, the correlation function of $c-$fermions is a product of correlation functions of the two orthogonal sectors
\begin{eqnarray}
\begin{aligned}
\label{eq.correlation_f}
\langle c_i^{\dagger}c_{j}\rangle=\langle f_i^{\dagger}f_j\rangle\langle \tilde{\tau}_i^x\tilde{\tau}_j^x\rangle,
\end{aligned}
\end{eqnarray}
which is the defining feature of the OM, namely the correlation function of physical charges vanishes for $\langle \tilde{\tau}_i^x\tilde{\tau}_j^x\rangle=0$, despite non-zero correlations of the $f-$fermion sector~\cite{nandkishore2012orthogonal}. Note, for $i=j$, local density for $c-$ and $f-$ fermions are identical and, thus, the kinetic constraints are the same. 

In the following, we will analyze the quench dynamics from a special initial state, and confirming that our non-integrable model can host persistent oscillations, meanwhile generating volume-law entanglement.

\textit{Coherent oscillation with volume-law entanglement.--}
\begin{figure}
	\centering
	\includegraphics[width=0.99\linewidth]{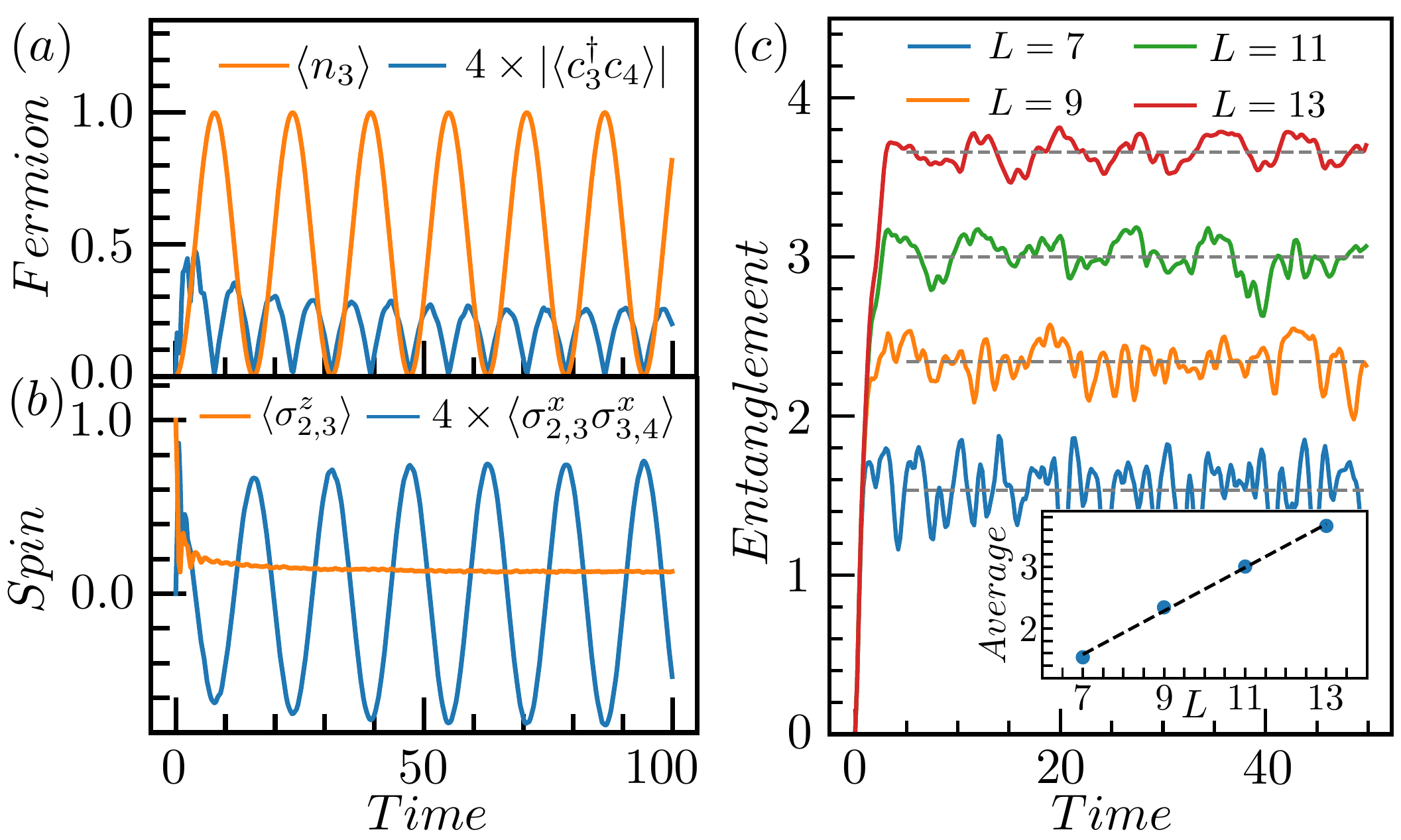}
	\caption{Quench dynamics of local observables (orange) and nearest neighbor correlation functions (blue), in panel (a) for physical $c-$fermions and panel (b) for $\sigma-$spins ($L=20, h=1, J=0.7, t=0.2, g_z=-0.4$). (c) Rapid increase of the half-system von Neumann entanglement to a volume-law plateau. (inset) Average entropy for the plateau as a function of system size.}
	\label{fig:dynamics}
\end{figure}
We consider the tensor-product of $c-$fermions and $\sigma-$spins as  $|0\rangle=|\psi\rangle_c \otimes|S\rangle_{\sigma}$, where $|S\rangle_{\sigma}=\ket{\uparrow\uparrow\dots\uparrow}_{\sigma}$ and $|\psi\rangle_c$ represents the fermionic Fock state. For simplicity, we focus on a chain of length $L$ with three fermions, and the initial state involves one building block $\ket{010110}_c$ and empty sites $\ket{0\dots 0}_c$ of length $L-6$, i.e., $|\psi\rangle_c = \ket{0101100\dots0}$
with periodic boundary conditions. The qualitative results are unaltered for a finite fermion density as long as the kinetic constraints lead to isolated acive sites, e.g. here sites 3 and 4, separated by frozen segements.

Fig.~\ref{fig:dynamics} depicts the dynamics of local observables (orange) and nearest neighbor correlations (blue) for physical $c-$fermions and $\sigma-$spins in panel (a) and (b), respectively. By construction, the fermionic occupation $\langle n_i\rangle$ exhibits persistent coherent oscillation only for $i=3,4$ and remains constant on the other frozen sites (not shown). The oscillation can be analytically obtained by projecting the Hamiltonian $H_f$ to the basis $\ket{01}_f,\ket{10}_f$ including only the active sites
 as $H_f=-t(f_3^{\dagger}f_4+f_4^{\dagger}f_3)$. The initial state $\ket{01}_f$ hence starts oscillating coherently within the isolated subspace, resulting in the occupation $\langle n_3\rangle(s)=\sin^2(ts)$ where $s$ stands for time, which oscillates with frequency $2t$ .
 The nearest-neighbor correlation function $\langle c_3^{\dagger}c_{4}\rangle$  oscillates as well because our choice of quench parameters leads to $\langle \tilde{\tau}_i^x\tilde{\tau}_{i+1}^x\rangle\neq0$.

The local magnetization in Fig.~\ref{fig:dynamics} (b) quickly equilibrates, but surprisingly the correlation between its adjacent spins $\langle \sigma^x_{2,3}\sigma^x_{3,4}\rangle$ has persistent oscillations. Indeed, according to Eq.~\eqref{eq.spin}, it synchronizes with the local fermionic dynamics as long as $\langle \tilde{\tau}_i^z\rangle$ does not vanish, which happens for quenches within the disordered phase of $H_{\tilde{\tau}}$.

The entanglement behavior of the composite model strongly depends on the choice of partition of the system. 
In Fig.~\ref{fig:dynamics} (c), we show the half-system entanglement $S_{L/2}=-\mathrm{Tr}\left[\rho_{L/2}\log\rho_{L/2}\right]$ where $\rho_{L/2}$ represents the reduced density matrix for a contiguous half of the system defined by the physical d.o.f. of $c-$fermions and $\sigma-$spins. A rapid entanglement growth is observed which quickly  saturates around a plateau. The average entropy of the plateau is plotted as a grey dashed line and its value scales linearly with system size as seen in the inset of the figure, confirming the volume-law behavior of the entanglement. 
In terms of the separable d.o.f., the non-intergrable $\tilde{\tau}-$spins are also volume-law entangled, whereas the $f-$fermions are only area-law entangled as the spreading of correlation is prohibited by frozen sites. Indeed,
the kinetic constraints retain the exact separation of active and frozen building blocks, thus only area-law entanglement saturation can be achieved also for $c-$fermions. This peculiar behavior of entanglement between different components is precisely the defining feature of the QDLs but here with the additional novel feature of persistent oscillations. 
\begin{figure}
	\centering
	\includegraphics[width=0.99\linewidth]{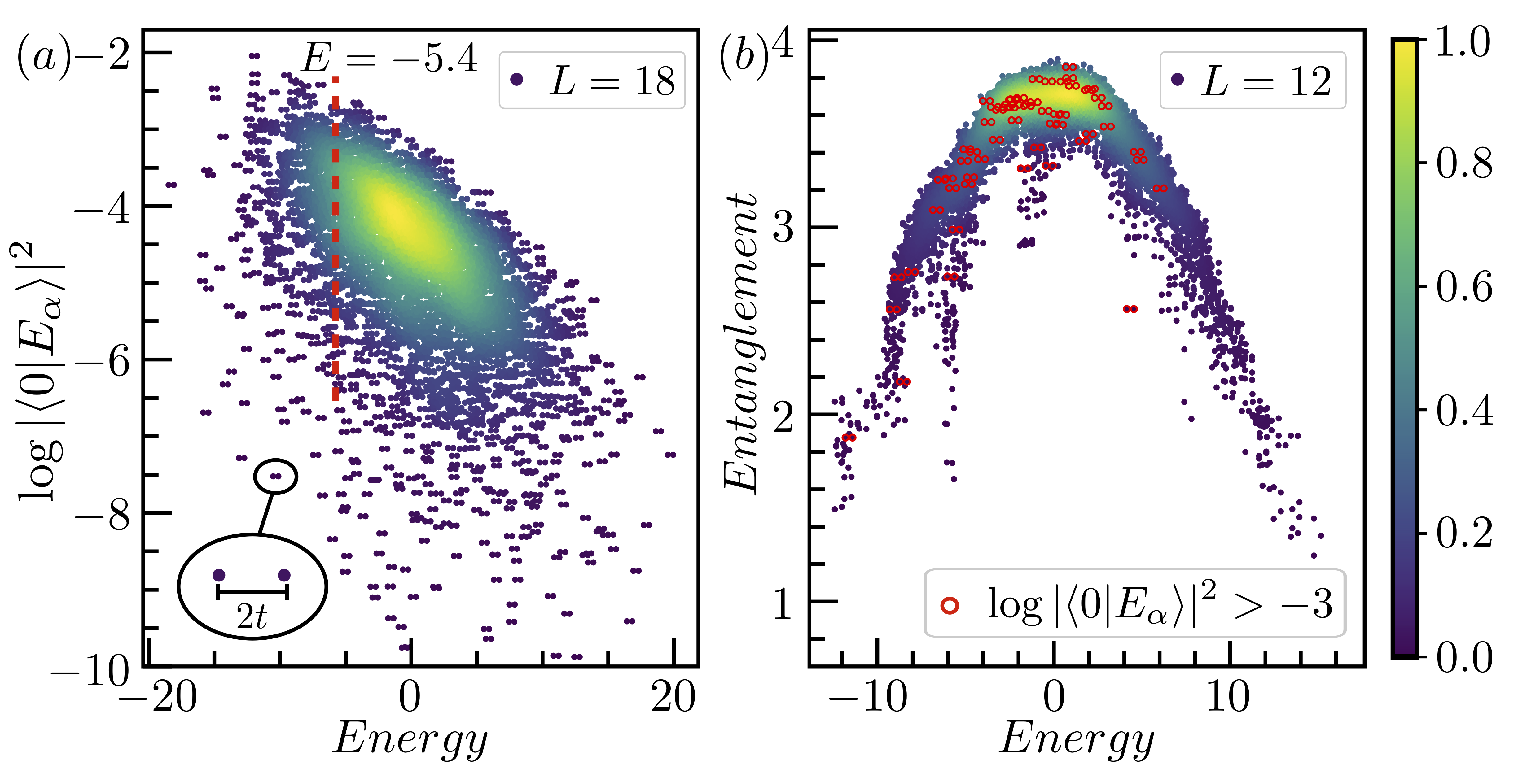}
	\caption{(a) Overlap between initial state and eigenstates for $L=18$. All states are paired with energy difference $2t$. Energy of the initial state is depicted as a red line. (b) Half-system von Neumann entanglement entropy for each eigenstate for $L=12$. Eigenstates with large overlap with initial state are circled in red. The width of the distribution of energy reduces due to a smaller system size ($h=1 ,J=0.7, t=0.2, g_z=-0.4$).}
	\label{fig:spectrum}
\end{figure}

\textit{Spectrum.--}The peculiar features of {\it orthogonal} quantum many-body scars can also be understood via properties of the many-body spectrum. 
Fig.3 (a) depicts the overlap $|\langle 0|E_{\alpha}\rangle|^2$ between the initial state $\ket{0}$ and eigenstates $\ket{E_{\alpha}}$ on a log scale, versus the eigenenergy $E_{\alpha}$. 
The initial state has overlap with an extensive number of eigenstates. The dotted red line indicates the energy of the initial state $E=\langle 0|H|0\rangle=-5.4$ which locates in the bulk of the spectrum and, therefore, without the kinetic constraints generic ETH behaviour would be expected. However, it does not happen here as seen in Fig.~\ref{fig:dynamics}, thus suggesting the violation of ETH which can be understood by decomposition:
eigenvalues for the physical model can be obtained as $E_{\alpha}=E_{\tilde{\tau}}+ E_{f}$, where $E_{\tilde{\tau}/f}$ denotes eigenvalues for $\tilde{\tau}-$spins and $f-$fermions respectively. As $E_{f}$ only takes value $\pm t$ in the reduced subspace, all eigenstates shown in Fig.~\ref{fig:spectrum}(a) are paired with another one with the same overlap but with energy separation  $2t$. 
These eigenstate pairs have different expectation values for the fermion number and the energy separation $2t$ is directly related to the frequency of persistent oscillations. 

Although ETH is strongly violated, entanglement can still be largely generated for most of the eigenstates through the background spins. 
Fig.~\ref{fig:spectrum}(b) depicts
the entanglement entropy for the half-system $S_{L/2}$ for all eigenstates in the reduced Hilbert space~\cite{appendix}, which increases towards the bulk of the spectrum. Note that the distribution of energies is different from Fig.~\ref{fig:spectrum}(a) due to a smaller system size, but in both cases the energy of initial state locates in the bulk of the spectrum. The states with dominant overlap with the initial state ($\log|\langle 0|E_{\alpha}\rangle|^2>{-3}$) are highlighted in red, and most of them are highly entangled with $S_{L/2}\gtrsim 3.5$.
This is in sharp contrast to known quantum many-body scars, which normally exhibit exceptionally low entropy in accordance with their non-ergodic dynamical behavior.

\begin{figure}
		\includegraphics[width=0.99\linewidth]{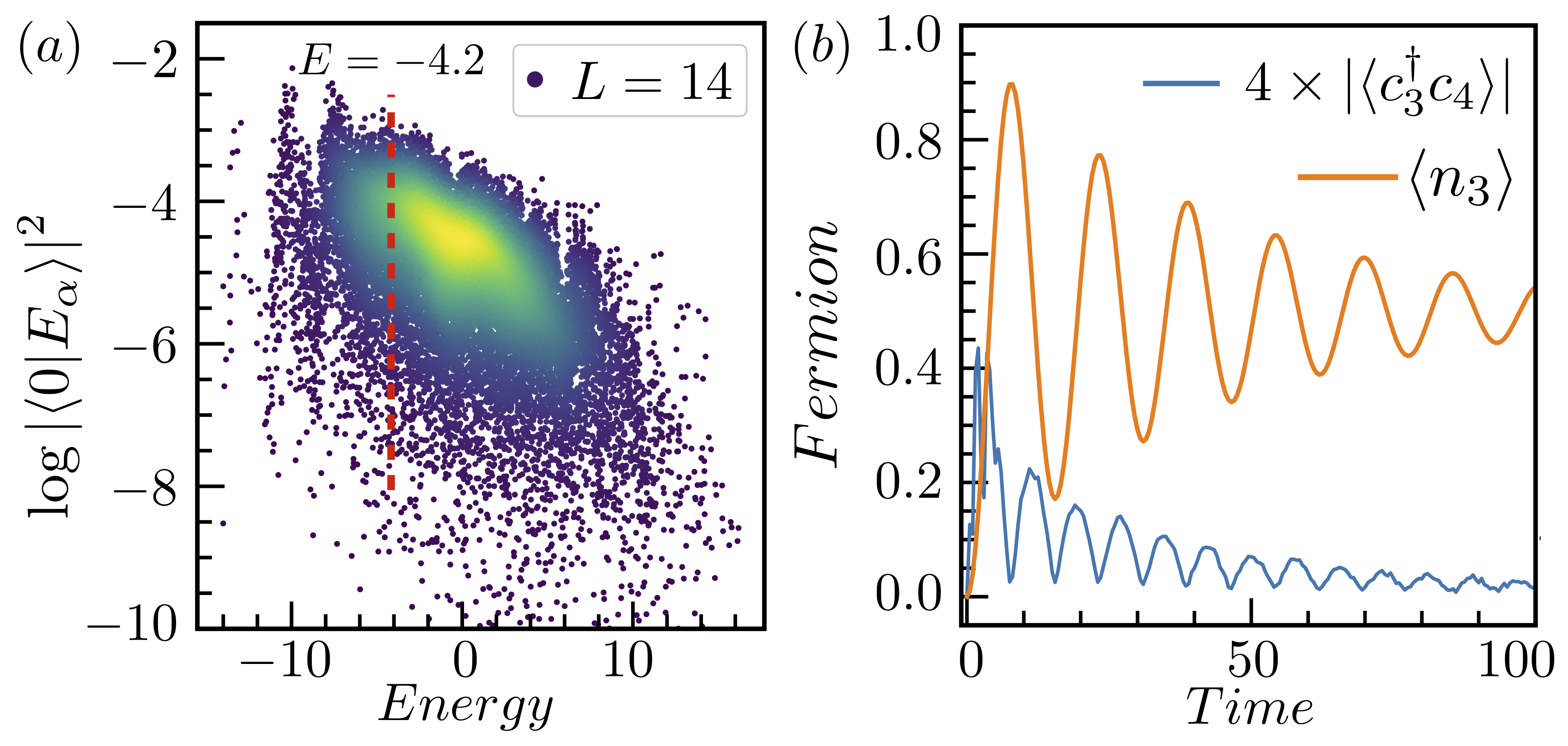}
	\caption{(a) Overlap between our initial state and eigenstates $E_{\alpha}$. (b) Damping of the local dynamics and correlation function. Parameters are chosen as $L=14, J=0.7, h=1, t=0.2, g_z=-0.4$ with an additional interaction $g_x=0.1$ which breaks the fine-tuned construction and separability into to independent sectors.}
	\label{fig:breakingMapping}
\end{figure}
\textit{Separability breaking.--}
It might seem that our results hinge on the fine-tuned construction of Eq.~\eqref{eq.original},
but they remain valid in the presence 
of generic perturbations which break the separability of the Hamiltonian into separate terms $H_f$ and $H_{\tilde{\tau}}$.
This can be demonstrated with the example of the perturbation
$g_x\sum_i\sigma_{i-1, i}^x\sigma_{i, i+1}^x$
which is lacking the factor $(-1)^{c_i^{\dagger}c_i}$ as in the original model Eq.~\eqref{eq.original}. As such, although the mobility of $c-$fermions remains restricted to two active sites, their dynamics cannot be separated anymore from the background $\sigma-$spins. These spins can then act as a thermalizing bath for the fermions.

The overlap between initial states and eigenstates $|\langle 0|E_{\alpha}\rangle|^2$ is shown in Fig.~\ref{fig:breakingMapping} (a), and the mean energy $E=-4.2$ is still close to the middle of the spectrum. The typical eigenstate pairing as seen in Fig.~\ref{fig:spectrum} is absent as the perturbation violates the separability. Fig.~\ref{fig:breakingMapping}(b) depicts the dynamics of local observables (orange) and correlation functions (blue) of $c-$fermions for perturbation strength $g_x=0.1$. The oscillations of the fermionic occupation is not infinitely long-lived anymore but decays with a finite lifetime, which scales as $t_{th}\sim g_x^{-2}$ as predicted by Fermi's Golden rule \cite{appendix,crowley2021partial},
eventually saturating to a constant value suggesting thermalization. The same behavior also appears for correlation functions, which decay towards zero at long times. Thus, when perturbing away from the fine-tuned construction our {\it perfect} orthogonal scars turn into orthogonal scars with long but not persistent oscillations.

\textit{Disucssion.--}
We have presented a simple model of orthogonal quantum many-body scars by combining kinetic constraints with the fractionaliztion mechanism of the orthogonal metal, which demonstrates that non-ergodic dynamics can coexist with a rapid volume-law entanglement entropy generation in a standard quench protocol.  

The key elements of our model are potentially realisable in experiments. For example, the density-dependent tunneling can be obtained in cold atomic gases via modulating external magnetic fields periodically in the vicinity of Feshbach resonances~\cite{meinert2016floquet,kotochigova2014controlling}, or in the presence of dominant nearest-neighbor density-density interaction~\cite{de2019dynamics}. The spin background minimally coupled to the fermions can be realized by exploiting the tool boxes in development for lattice gauge theory simulations~\cite{banuls2020simulating,martinez2016real,schweizer2019floquet,yang2020observation}. 

Our results enrich the phenomenology of quantum many body scars and  of quantum disentangled liquids. It will be worthwhile to explore whether the orthogonal scars of our model can appear in other correlated multi-component systems. Similarly to the fractionalization of the OM, which can emerge as a low energy feature captured by slave-particle descriptions of more generic systems~\cite{nandkishore2012orthogonal}, also the kinetic constraints can emerge as effective interactions. Hence, potential candidates for realising orthogonal quantum many-body scars are for example  multi-orbital Hubbard systems and the possibility to alter their dynamics via Floquet engineering opens another promising pathway. 

The one-dimensional example presented here can be readily generalized to higher dimensions by combining density assisted tunneling with the two-dimensional OM construction which is considerably richer because the background spin sector can be topologically ordered ~\cite{nandkishore2012orthogonal}. An alternative direction for future research is to explore the interplay between kinetic constraints and local gauge symmetry. For example, combining the disorder-free localization mechanism~\cite{smith2017disorder,smith2018dynamical,brenes2018many,karpov2020disorder,paulson2020towards,papaefstathiou2020disorder} with kinetic constraints could potentially lead to non-ergodic gauge field dynamics. 

In conclusion, we expect that the interplay of fractionalization and kinetic constraints in multi-component systems will lead to more surprises and unexpected non-equilibrium physics. 

{\it Acknowledgements.--}
We acknowledge helpful discussion with Andrea Pizzi and Markus Heyl. H.Z. was supported by a Doctoral-Program Fellowship of the German Academic Exchange Service (DAAD). A.S. was supported by a Research Fellowship from the Royal Commission for the Exhibition of 1851. We acknowledge support from the Imperial-TUM flagship partnership.

\bibliography{OM_ref} 
\newpage
\appendix

\section{Numerical details}
All of the dynamics and the specturm are obtained by Exact Diagonalization(ED) by Python package QuSpin~\cite{weinberg2017quspin,weinberg2019quspin}. As analyzed in the main content, the density-dependent tunneling for $c-$fermion significantly restricts its dynamics. Therefore, one can employ such a feature to enlarge the system size simulatable by ED with reduced fermionic basis, even when the separability is broken in the presence of non-vanishing $g_x$ as in Fig. 4. For instance, the basis for $c-$fermion with three particles and $L$ sites indeed only involves two Fock states as $|\psi\rangle_c = \ket{0101100\dots0}$ and $|\psi\rangle_c = \ket{0110100\dots0}$, since there are only two active sites contributing to the dynamics. The spectrum is also obtained by diagonalizing the Hamiltonian in the reduced basis.

\section{Lifetime of orthogonal scars}
\begin{figure}[h]
	\centering
	\includegraphics[width=0.95\linewidth]{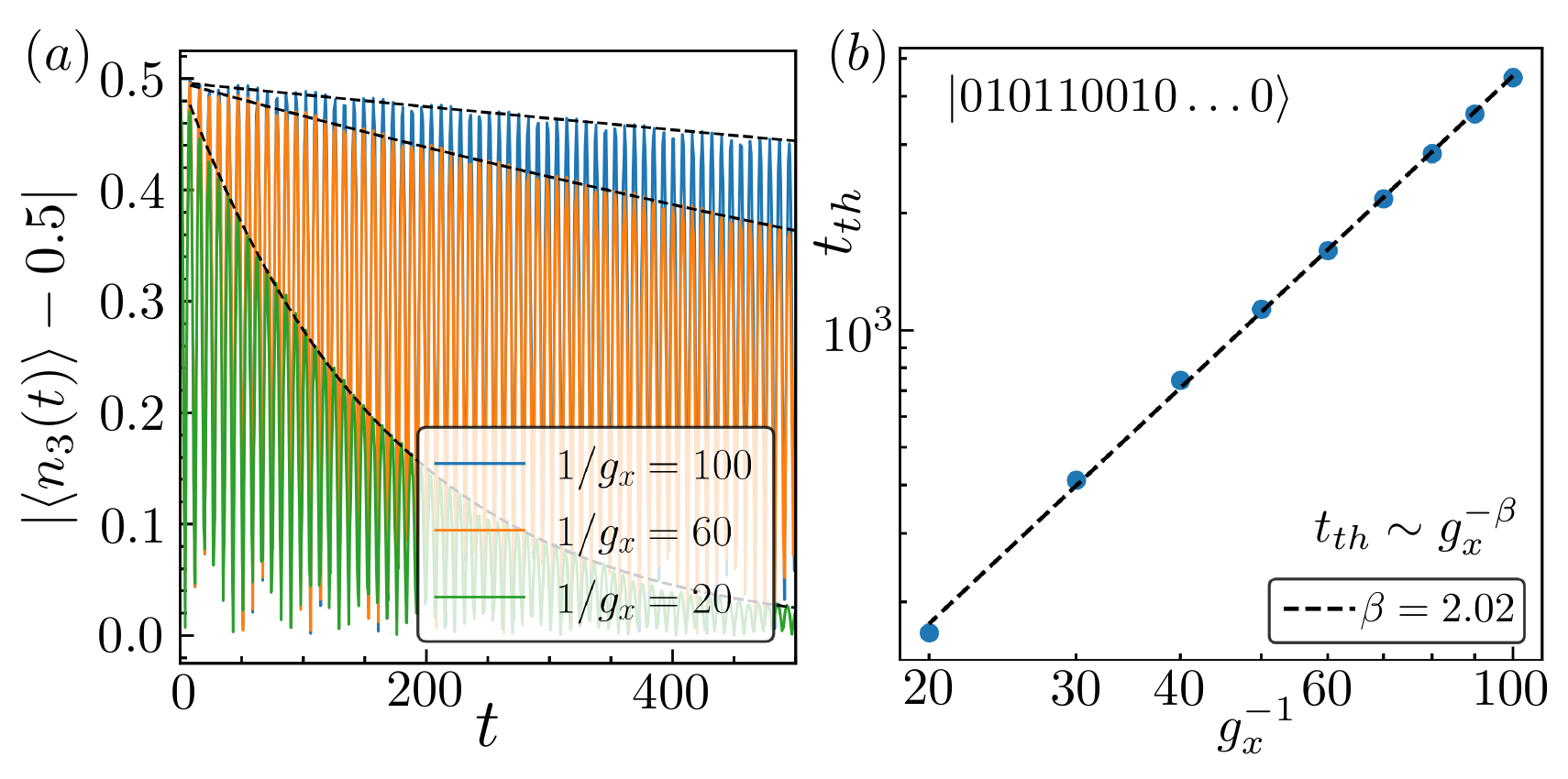}
	\caption{(a)Damped oscillations of the local fermionic dynamics. (b)Lifetime $t_{th}$ of the oscillation versus the separability breaking perturbation $g_x$, which fits well with $t_{th}\sim g^{-2}$ ($h=1 ,J=0.7, t=0.2, g_z=-0.4$).}
	\label{fig:thermaldynamics}
\end{figure}

The perturbation $g_x\sum_i\sigma_{i-1, i}^x\sigma_{i, i+1}^x$ will break the separability of the model, hence the coherent oscillation has a finite lifetime. To quantify the lifetime, in panel (a) of Fig. \ref{fig:thermaldynamics}, we depict the deviation between the occupation number and its thermal equilibration starting from the initial state $\ket{\uparrow\uparrow\dots\uparrow}\otimes\ket{010110010\dots 0}$ for $L=14.$ The oscillation is damping and we fit the envelop as an exponential function $0.5e^{-t/t_{th}}$ as indicated as the black dashed curve. In panel (b), the lifetime $t_{th}$ is plotted versus the inverse of the separability breaking perturbation $g_x^{-1}$ which scales approximately as $t_{th}\sim g_x^{-2}$ as predicted by the Fermi's Golden rule \cite{crowley2021partial}.

\end{document}